\documentstyle[hhline,natbib2,natbibmnfix,epsfig,lscape]{mn}
\bibpunct[,]{(}{)}{;}{a}{}{}

\newif\ifAMStwofonts

\def\lcdm{$\Lambda$CDM }

\def\Bband{$B$-band }

\newcommand{\hkpc}{\mbox{$h^{-1}$ kpc} }

\newcommand{\hMpc}{\mbox{$h^{-1}$ Mpc} }

\newcommand{\hMpcInvThreeKet}{\mbox{$h^{3} \tx{Mpc}^{-3}$)} }

\newcommand{\hMpcCom}{\mbox{$h^{-1}$ Mpc,} }
\newcommand{\hMpcDot}{\mbox{$h^{-1}$ Mpc.} }

\newcommand{\hmsun}{\mbox{$h^{-1}$ $M_{\odot}$} }

\newcommand{\kms}{\mbox{km s$^{-1}$} }

\newcommand{\kmsDot}{\mbox{km s$^{-1}$.} }

\newcommand{\kmsKC}{\mbox{km s$^{-1}$),} }

\newcommand{\msun}{\mbox{$M_{\odot}$} }

\newcommand{\Mvir}{\mbox{$M_{\rmn{200}}$} }

\newcommand{\Vvir}{\mbox{$V_{\rmn{200}}$} }

\newcommand{\IRAS} {\emph{IRAS}\ }

\newcommand{\PSCZ} {\emph{PSCz}\ }

\newcommand{\CfA} {\emph{CfA}\ }

\newcommand{\SSRS} {\emph{SSRS}\ }

\newcommand{\LCRS} {\emph{LCRS}\ }

\newcommand{\ORSCom} {\emph{ORS,}\ }

\def\la{\mathrel{\hbox{\rlap{\hbox{\lower4pt\hbox{$\sim$}}}\hbox{$<$}}}}
\def\ga{\mathrel{\hbox{\rlap{\hbox{\lower4pt\hbox{$\sim$}}}\hbox{$>$}}}}

\newcommand{\bc}{\begin{center}}
\newcommand{\ec}{\end{center}}

\newcommand{\tx}[1] {\rmn{#1}}

\newcommand{\refg} {$Ref$ }		
\newcommand{\refgDot} {$Ref$. }		
\newcommand{\refgCom} {$Ref$, }		
\newcommand{\GL}[1] {$GL_{#1}$}		
\newcommand{\HM}[1] {$HM_{#1}$}		
\newcommand{\GC}[1] {$GC_{#1}$}		
\newcommand{\GM}[1] {$GM_{#1}$}		
\newcommand{\LFM}[1] {$LFM_{#1}$}		
\newcommand{\LFV}[1] {$LFV_{#1}$}		
\newcommand{\V} {$V$ }		
		
\newcommand{\VCom} {$V$, }

\ifoldfss
  \newcommand{\rmn}[1] {{\rm #1}}

  \ifCUPmtlplainloaded \else
    \NewTextAlphabet{textbfit} {cmbxti10} {}
    \NewTextAlphabet{textbfss} {cmssbx10} {}
    \NewMathAlphabet{mathbfit} {cmbxti10} {} 
    \NewMathAlphabet{mathbfss} {cmssbx10} {} 
  \fi
  \ifAMStwofonts
    \ifCUPmtlplainloaded \else
      \NewSymbolFont{upmath} {eurm10}
      \NewSymbolFont{AMSa} {msam10}
      \NewMathSymbol{\upi}     {0}{upmath}{19}
      \NewMathSymbol{\umu}     {0}{upmath}{16}
      \NewMathSymbol{\upartial}{0}{upmath}{40}
      \NewMathSymbol{\leqslant}{3}{AMSa}{36}
      \NewMathSymbol{\geqslant}{3}{AMSa}{3E}

      \let\leq=\leqslant \let\le=\leqslant
      \let\geq=\geqslant 
    \fi
  \fi
\fi 

\ifnfssone
  \newmathalphabet{\mathit}
  \addtoversion{normal}{\mathit}{cmr}{m}{it}
  \addtoversion{bold}{\mathit}{cmr}{bx}{it}
  \newcommand{\rmn}[1] {\mathrm{#1}}

  \newmathalphabet{\mathbfit} 
  \addtoversion{normal}{\mathbfit}{cmr}{bx}{it}
  \addtoversion{bold}{\mathbfit}{cmr}{bx}{it}
  \newmathalphabet{\mathbfss} 
  \addtoversion{normal}{\mathbfss}{cmss}{bx}{n}
  \addtoversion{bold}{\mathbfss}{cmss}{bx}{n}
  \ifAMStwofonts
    \ifCUPmtlplainloaded \else
      \UseAMStwoboldmath
      \makeatletter
      \new@mathgroup\upmath@group
      \define@mathgroup\mv@normal\upmath@group{eur}{m}{n}
      \define@mathgroup\mv@bold\upmath@group{eur}{b}{n}
      \edef\UPM{\hexnumber\upmath@group}
      \new@mathgroup\amsa@group
      \define@mathgroup\mv@normal\amsa@group{msa}{m}{n}
      \define@mathgroup\mv@bold\amsa@group{msa}{m}{n}
      \edef\AMSa{\hexnumber\amsa@group}
      \makeatother
      \mathchardef\upi="0\UPM19
      \mathchardef\umu="0\UPM16
      \mathchardef\upartial="0\UPM40
      \mathchardef\leqslant="3\AMSa36
      \mathchardef\geqslant="3\AMSa3E

      \let\leq=\leqslant \let\le=\leqslant
      \let\geq=\geqslant 
    \fi
  \fi
\fi 

\ifnfsstwo
  \newcommand{\rmn}[1] {\mathrm{#1}}

  \DeclareMathAlphabet{\mathbfit}{OT1}{cmr}{bx}{it}
  \SetMathAlphabet\mathbfit{bold}{OT1}{cmr}{bx}{it}
  \DeclareMathAlphabet{\mathbfss}{OT1}{cmss}{bx}{n}
  \SetMathAlphabet\mathbfss{bold}{OT1}{cmss}{bx}{n}
  \ifAMStwofonts
    \ifCUPmtlplainloaded \else
      \DeclareSymbolFont{UPM}{U}{eur}{m}{n}
      \SetSymbolFont{UPM}{bold}{U}{eur}{b}{n}
      \DeclareSymbolFont{AMSa}{U}{msa}{m}{n}
      \DeclareMathSymbol{\upi}{0}{UPM}{"19}
      \DeclareMathSymbol{\umu}{0}{UPM}{"16}
      \DeclareMathSymbol{\upartial}{0}{UPM}{"40}
      \DeclareMathSymbol{\leqslant}{3}{AMSa}{"36}
      \DeclareMathSymbol{\geqslant}{3}{AMSa}{"3E}

      \let\leq=\leqslant \let\le=\leqslant
      \let\geq=\geqslant 
    \fi
  \fi
\fi 

\ifCUPmtlplainloaded \else
  \ifAMStwofonts \else 
    \def\upi{\pi}
    \def\umu{\mu}
    \def\upartial{\partial}
  \fi
\fi
	
\title{Voids in the Simulated Local Universe}
\author[H. Mathis and S. D. M. White]
	{H.~Mathis$^1$\thanks{Email: hmathis@mpa-garching.mpg.de} and
	S.~D.~M.~White$^1$ 	
	\\	
        $^1$Max--Planck--Institute for Astrophysics, D-85741 Garching, Germany}

\pagerange{\pageref{firstpage}--\pageref{lastpage}}

\begin{document}

\maketitle

\label{firstpage}


\begin{abstract}

We use simulations of the formation and evolution of the galaxy
population in the Local Universe to address the
issue of whether the standard theoretical model succeeds
in producing empty regions as large and as dark as the observed
nearby ones. We follow the formation of galaxies in a \lcdm universe 
and work with mock catalogues which can resolve the morphology of LMC
sized galaxies, and the luminosity of objects 6 times fainter. We 
look for a void signature in sets of virialized haloes selected by
mass, as well as in
 mock galaxy samples selected according to observationally
relevant quantities, like luminosity, colour, or morphology. 
We find several void regions with diameter $10 \hMpc$ in the
simulation where gravity seems to have swept away even the smallest
haloes we were able to track. We probe the environment density of
the various populations and compute luminosity functions for
galaxies residing in underdense, mean density and overdense
regions. We also use nearest neighbour statistics to check 
possible void populations, taking $L_{*}$ 
spirals as reference neighbours. Down to our resolution limits, 
we find that all types of galaxies 
avoid the same regions, and that no class appears to populate the voids 
defined by the bright galaxies.
 
\end{abstract}


\begin{keywords}
large--scale structure of the Universe -- galaxies: statistics --
galaxies: formation
\end{keywords}


\section[]{Introduction}
\label{sec:Intro}

Early concerns about the existence of a 
hypothetical, spatially homogeneous population
of galaxies may be traced back to the negative conclusions of 
\citet{So77},  well before it was realized that 
the local universe contains huge regions 
apparently empty of normal, optically selected
galaxies \citep{Ki81}. More recent observations targetted
specific classes of galaxy (e.g. dwarfs,
low surface brightness or star-forming galaxies) finding that the
optically-defined voids also seem strongly underdense in all
these objects (see Section~2 of \citealt{Pee01} for a list of
observations). There have been reports of the detection of some
galaxies in previously defined 
voids, like Bo\"{o}tes \citep{Dey90,Szo96a,Szo96b}, but these are
mostly normal, late-type spirals near the edge of the void region. 
Starting from a morphology density relation like that of
\citet{Dr80}, one might expect the
voids to be populated by a population of dwarf, faint
galaxies. However, even these populations are not observed. As noted by
\citet{Ki81} and underlined by \citet[hereafter P01]{Pee01}, \emph{if} there is still
substantial mass in the voids, the galaxies associated with
 DM haloes in the voids must be several magnitudes fainter than
$L_{*}$. The Magellanic-type irregulars close to the Local Group
\citep{Tu88,Pee89,Pee00b} seem to have formed in conditions quite
comparable to those occurring in the voids so one might expect to
find the same type of galaxy there.

Given that the same large underdense regions are observed in different surveys and
populations, several groups have recently focused on the precise determination
of the sizes and shapes of voids in large redshift surveys, like the \CfA
redshift survey \citep{Vo94a}, the \LCRS \citep{Mu00,Mu01}, the
\SSRS and \IRAS surveys \citep{ElAd97a,ElAd97b}, and the \PSCZ
survey \citep{Pl01}. When comparing the voids in the \IRAS 1.2 Jy
survey and in the \ORSCom \citet{ElAd00} found similar sizes and shapes,
although the galaxy populations are very different in the two samples.

Some authors \citep{Wh87ApJ,Vo94a,Mu00,Pl01} have used
 simulations of the dark matter distribution in CDM models to compare to
observations of ``voids''. All of this work identified galaxy sites by simple
statistical bias models rather than by \emph{ab
initio} galaxy formation modelling. Many authors also studied regions too small to
 provide reliable statistics. 
\citet{Vo94a} found that biased models (e.g. a flat, low-density CDM
 cosmology) generally produce voids that are
similar to the observations in the \CfA survey when considering bright
galaxies, but which are too empty when including fainter galaxies. \citet{Mu00}
found their void statistics to be more sensitive to the criterion for
galaxy identification than to the cosmological
parameters themselves. \citet{Pl01} compared the void shapes and sizes in the
\PSCZ with six different CDM models. They found the distribution of
void sizes to differ strongly between models. Models with high
$\sigma_{8}$ like OCDM and \lcdm provided the best fits.

The main drawback of these kinds 
of simulations is the necessity 
to put in the galaxies ``by hand'', using an \emph{ad hoc}
bias recipe, and/or by assigning morphologies using 
the observed morphology-density relation.

In recent years, numerical simulations of structure formation in cosmological volumes 
have been extended in two different ways to include an explicit
treatment of galaxy formation by means of
semi-analytic techniques. In the faster, but less direct technique,
galaxies are inserted in each dark halo at the time of observation
based on a Monte-Carlo realization of the merging and galaxy formation
history of a ``random'' halo of similar mass
\citep{Kau97,Gov98,Ben01a,Ben01b}. In the more direct simulation
technique, the full merging history of each halo in a simulation is
stored and the semi-analytic formation prescriptions are implemented
 on these histories, producing a simulation in which the detailed
formation history of each galaxy is followed explicitly 
\citep{Kau99a,Kau99b,Dia99,Dia01,Som01,Spr01,Ma01}.

The simulations of \citet{Kau97} were able to resolve LMC-like
galaxies. These authors considered the void probability
function (VPF) in an Einstein--de Sitter universe. They computed the VPF for 
bright ($M_{B}<-20$) and faint ($M_{B}<-18$) mock galaxies and for two
corresponding random sets of DM particles in the simulation, each
chosen to have the same number density as the selected galaxies. While they found in
their Figure~14 a similar VPF for bright galaxies and the associated
dark matter set, the VPF of their faint galaxies exhibited 
an extended tail as compared to the DM, showing
that these faint objects tend to respect the voids defined by the
bright ones, and to segregate from the dark matter.

\citet[hereafter K99]{Kau99} (see also \citealt{Ben01a}) had coarser
resolution, and did not
compute void statistics. Instead one can get an idea of their
predicted voids by means of the pictures of the galaxy distribution they provide. 
As noted by P01, visual inspection of the simulations shows large
regions between the concentrations of $L_{*}$ 
galaxies in clusters and filaments, which appear almost empty of
galaxies (see figure~5 of K99). However, the extent to which these
simulated voids retain 
a significant fraction of the mass, in the
form of LMC-sized or smaller DM haloes, is unclear, even in the low-density \lcdm
model. If there are haloes in the voids, then no galaxies have been
associated with them by the galaxy formation algorithm, and one has to check if
this is due to the resolution limit or to a well-determined physical process.

\emph{If} simulations still show a substantial number of lower mass 
DM haloes in ``voids'', 
P01 considers this observed suppression of galaxy formation 
a ``potential crisis'' for the \lcdm paradigm. Note, however, that
there may be possible remedies: for example, ionizing fluxes from the first
structures may suppress nearby dwarf galaxy formation
(\citealt{Re85,Sri97,Cen00}, see also \citealt{Fr00}). Alternatively, 
the Warm Dark Matter model of \citet{Bo00} may
help to solve the issue.

The goal of this paper is to help to clarify the situation of voids from a \emph{numerical}
point of view, using the \lcdm simulation presented in \citet[hereafter M01]{Ma01}. This
simulation was constrained so that the present-day DM
distribution mimics the large scale structure
of the local universe up to a distance of 8000 \kms from the Milky-Way. 
Galaxy formation was followed in this simulation using techniques
similar to those of K99 and mock catalogues were extracted for
comparison to the observed nearby galaxy distribution. The main
clusters in the simulation appear at the position of observed clusters
like Coma, Virgo, Hydra, Perseus and Centaurus and larger scale
structures, including voids, correspond well. The simulations can track 
the formation and morphological evolution of dwarf galaxies of 
the size of the LMC, and can resolve the luminosity of
galaxies 6 times fainter. The corresponding \emph{morphology} and \emph{luminosity}   
resolution limits of the simulation (defined as the mean, 
present-day \Bband luminosities of the central galaxy of a 10 and 100
particle halo respectively) are $M_{B}=-16.27$ and $M_{B}=-18.46$. 
They correspond to halo virial circular velocities of haloes of $\sim 54$ and
$\sim 103$ \kms and masses of $\sim 3\times 10^{10}$ and $\sim 3\times
10^{11} \hmsun$ respectively (here and below we take h=0.7 when
quoting absolute magnitudes).

The morphology resolution of the CR \lcdm simulation is  
only a factor 3 better than that of
\citet{Kau97} but the volume simulated is 3 times larger. Also, the
 \lcdm model currently seems a better bet than the EdS model they
 assumed. In addition the current galaxy formation algorithm is based on  
a much more detailed model including morphological evolution of the galaxies via
merging, explosive (bursts) and quiescent star formation, and feedback. 

Below we begin by using simple, ``one-point'' statistics 
to bring out a hypothetical void 
population. From the DM skeleton alone, 
we are able to check the claim 
of P01 on the fraction and typical masses of 
the haloes residing in underdense volumes. Simply
by visual inspection of the DM distribution, 
we highlight regions which are depleted of even the smallest 
haloes that we can follow.

Then, within the whole simulated galaxy population, we select several 
subsamples (candidates for a void population), splitting according to
a variety of possibly relevant properties: luminosity, colour,
morphology. Also, we extract sets of haloes, binned by 
total mass. We study the typical densities 
(estimated on a 5 \hMpc scale) 
 in which these various populations reside, with the goal of
finding a first signature of a void population. We recover the well-known
pattern of halo bias. Blue, star-forming
galaxies tend to reside in underdense regions, a
trend which is also present but is not 
so marked for bulge-less galaxies.

Next we estimate the environment density dependence of the galaxy
luminosity function, focussing on the variation of the shape and the
normalization of the LF with the local mass density. We compute LFs in
a series of density bins covering equal total volumes, and then
in bins containing equal total mass. While the overall normalization of the LF is
quite different between equal-volume bins, it is very similar between
equal-mass bins. We find some tendency for the faint-end 
slope of the LF to vary with DM environment density.

Finally, we consider ``multi-point'' probes like correlation functions
and nearest neighbour statistics. The correlation functions of our
samples provide a check of the selection procedure. We recover the
usual behaviour: in particular, red galaxies are much more clustered
than blue ones, in agreement with K99. We stress, however, that it is
difficult to deduce the relative spatial distribution of populations
on the basis of their correlation functions.

Since \citet{So77}, nearest neighbour statistics have been a common observational tool to
discriminate between spatial distributions of
objects. We carry out such an analysis in real space for our series 
of subsamples, taking $L_{*}$ spiral galaxies as the neighbour
population (following P01). We find that blue, star-forming galaxies
are more weakly associated with $L_{*}$ spirals than are
other $L_{*}$ spirals but otherwise we find no significant effect. 
We conclude that there is no dwarf-type void
population in the simulation, down to our resolution limit.

The disposition of our paper is as follows: in
Section~\ref{sec:Visual}, we begin with a qualitative, visual
comparison between the galaxies that have formed in 
a void and in a cluster environment. The distribution
of haloes in our void does not support the assertion of P01 that
the voids should contain small haloes.
Then, in Section~\ref{sec:Samples}, we describe the 
characteristics of our galaxy and DM halo subsamples. 
We show the distribution of our reference spirals in a slice
spanning the supergalactic plane. 
Section~\ref{sec:Envi} evaluates the environment 
densities for the various mock samples,
and computes the density dependence of LFs. 
We switch to two-point statistics in Section~\ref{sec:Clus}. We check 
the correlation functions of our subsamples and then carry out 
a nearest neighbour analysis similar to that of P01. We summarize 
and conclude in Section~\ref{sec:CCL}.


\section[]{An example of the galaxy distribution in a ``void''}
\label{sec:Visual}

Figure~3 of M01 (see also the bottom right panel of Fig.~\ref{fig:4PanelsFig} below) 
plots the simulated galaxy distribution within 8000 \kms of
the Milky-Way, in a slice of thickness 30 \hMpc centred on the
supergalactic plane. Two large voids are visible: the first at
(-60,-30) \hMpc in (SGX,SGY), with a diameter of $\sim 20 \: \hMpc$, the
other nearly opposite to the MW, at (40,50), where a galaxy separates
two smaller voids, of diameter  $\sim 10\: \hMpc$ each. 
The same voids are seen in the real galaxy distribution 
for example the \PSCZ survey
(consider Figure~2 of \citealt{Pl01}).

As an example, we focus here on the void located at
(40,50) \hMpcCom since it lies further inside high resolution region
of the simulation, and may be less
affected by the transition from the high-resolution to the
low-resolution zone. We excise a cubic region of side
24 \hMpc centred on the middle of the void.

The top left panel of Fig.~\ref{fig:4PanelsFig} shows the distribution
of all galaxies within this cube brighter than our luminosity resolution limit.
The colors of the symbols scale with the $B-V$ index, and their sizes
with the B-magnitude of the galaxies. Because of the galaxy 
formation scheme adopted in M01, all DM haloes 
in the simulation with 10 or more particles 
contain at least one galaxy. Furthermore, with our
definition of the luminosity resolution limit (the mean luminosity of the
central galaxy of a 10-particle halo), one expects that a fair fraction of
the haloes more massive than $10\times M_{part}=3.57\times10^{10}
\hmsun$ will have an associated galaxy with $M_{B}\le-16.27$, and so
will appear on the top left picture of
Fig.~\ref{fig:4PanelsFig}. In the region shown, 75\% 
of all haloes have a central galaxy at least this bright. 
This is shown explicitly in the bottom left panel of Fig.~\ref{fig:4PanelsFig}, 
where we project the same low density region as 
in the top left panel, and mark with circles the positions of
all DM haloes with 10 or more particles. 

There are two obvious voids completely depleted of haloes, with diameters of
order 10 and 8 \hMpc and depths of 24 \hMpcDot This contradicts 
P01 who claims that in a \lcdm model, low mass 
haloes should spread fairly evenly through the voids defined by
the larger ones. In fact, gravity seems to have moved them out of these
voids. 
 
\begin{figure*}		
\begin{minipage}{160mm}			
	\centering


\vspace{1cm}

	\caption{Top left: the present-day distribution of all galaxies brighter than
	the resolution limit ($M_{B}<-16.27$) in the void region
	defined in Section~\ref{sec:Visual} and selected from the \lcdm
	simulation of M01. The region shown is cubical and has a side
	of $24\:\hMpcDot$ The size of the symbols
	scales with the \Bband luminosity of the galaxies and their
	colour with the $B-V$ index. Top right: the present-day
	distribution of galaxies in a region surrounding the ``Virgo
	cluster'' in this same simulation. Galaxies are restricted to
	$M_{B}<-17$ (to avoid saturating the cluster region), colour
	and size of the circles scale as previously. Bottom left: 
	the halo distribution in the same region as in the
	top left picture. The same two voids as in
	galaxy distribution are apparent. Bottom
	right: the standard or ``reference'' late-type galaxies in the
	whole simulation, in a slice $30\:\hMpc$ thick in the SGZ
	direction encompassing the SG plane. The side of the
	picture is $180\:\hMpc$ long. Colour and size of the symbols
	 follow the previous rule. A high resolution copy of this
	Figure can be found at \bfseries{http://www.mpa-garching.mpg.de/NumCos/CR/Voids/}}
	\label{fig:4PanelsFig}

\end{minipage}
\end{figure*}

For comparison, we give in the top right
panel of Fig.~\ref{fig:4PanelsFig}
the galaxy distribution around the simulated Virgo cluster 
($\Mvir \sim 4\times10^{14}\hmsun$),
also plotted in a cubic region of side 24 \hMpc centred on the
cluster centre. The galaxy symbols follow the same rules and scales as in 
the left panel. Note that to avoid saturating the
cluster region, galaxies are plotted down to $M_{B}=-17$ only.


\section[]{Defining mock galaxy samples}
\label{sec:Samples}

The few isolated galaxies observed in the Bo\"{o}tes void are
 normal, late-type spirals \citep{Szo96a,Szo96b}. Nevertheless, 
there is a widespread belief that voids should be filled by late-type
 and low surface brightness dwarfs. Observational programs 
have targetted a number of different potential
``void'' populations:  \citet{Ed89} and \citet{Lee00} 
considered dwarf galaxies; \citet{Pu95} and \citet{Lin96} 
looked at blue compact galaxies (BCG); 
\citet{Sa90} separated their sample between high and low luminosity 
galaxies (with a limit at $M_{B}=-18$), and \citet{Bo93} looked for 
low surface brightness galaxies (LSB). In his unsuccessful  
quest for a possible void population in the \ORSCom P01
considered dwarfs/irregulars and LSBs, computing the distribution of
 distance from these objects to their nearest neighbour $L_{*}$ spiral.

To evaluate how biases arise in our simulations we will first study
the environments of dark haloes as a function of their mass, and
evaluate how their well-known clustering bias (e.g. \citealt{Mo96a})  
is echoed in the statistics that we later 
apply to our galaxy populations.

Then, using the photometric and morphological 
information provided by the simulation,
we study galaxy environments as a function of luminosity,
morphology, and colour. For example, we will
 compare the environments of faint galaxies to those of 
bright ones. To make contact with P01, we also analyze the nearest neighbour
statistics of our various halo and galaxy samples by looking for 
the nearest neighbour of each test object among a set of reference
$L_{*}$ spirals.

\subsection{Reference galaxies}
\label{sec:Samples:Ref}

These galaxies will be used as standard objects defining the
clustering of ``typical'' galaxies, in particular for our nearest
neighbour statistics. We select bright spirals within one magnitude 
of $M_{B,*}\sim -20.8$ (taken here as the typical magnitude of a Milky
Way look-alike): $-21.8 \la M_{B} \la -19.8$, and with Sa/Sb/Sc 
morphological type:  $1 < M_{\rmn{B,bulge}}-M_{\rmn{B,total}} < 2.2$
\citep{Si86,Vau91}. This criterion is similar to the one used in the 
construction of the Tully-Fisher relations in M01, but includes Sa
galaxies as well. We do \emph{not} require that these reference
galaxies be the central galaxies of their haloes however: 
they may also be satellites orbiting in large
clusters. We found 1471 such galaxies in the whole simulation, an
abundance very close ($7\times10^{-4} \hMpcInvThreeKet$ to the value
used by P01 for the spiral galaxies in his
analysis of the distribution of a population of LSB galaxies at 
distances $6000\leq cz \leq 9000\:\kmsDot$

The bottom right panel of Fig.~\ref{fig:4PanelsFig} shows these reference galaxies
in a slice of side-length $180\:\hMpc$ encompassing the supergalactic
plane. Even with  our applied morphological criterion, although the \emph{rms} spread 
in the colors of the reference galaxies is small, 
there are still some red galaxies. This is not unexpected, however; 
 most of these objects are ``anemic'' spirals in clusters. In the
following, we will refer to this sample of reference galaxies as \refgDot

\subsection{DM haloes}
\label{sec:Samples:Haloes}

We construct a series of dark matter halo samples binned 
according to their total mass (taken here to be the number of particles
 linked by the \emph{FOF} algorithm). We define our mass bins to span 
a factor 4, with the ``smallest bin'' having [10 40]
particles per halo and the largest [40960 163840] particles (there are
only 38 haloes in this last bin, and only 5 haloes with more than
163840 particles). We call the 7 resulting samples \HM{1} to \HM{7}
(Halo Mass) from the lightest to the heaviest. 
The \emph{lower} limits (in particle number and mass) 
for each of the 7 bins are given in the 
first 7 columns of the first two lines of Table~\ref{tab:BinLimits}, while
the last column gives the \emph{upper} limit of the most massive
bin, \HM{7}. The third line gives the number of haloes in
each bin.

\subsection{Luminosity selection}
\label{sec:Samples:Lum}

We define a series of luminosity bins to bring out the trends in the 
spatial distribution of objects as one goes from
bright galaxies down to dwarfs. Our most luminous bin (which
contains only 29 galaxies) has  $M_{B}<-22.5$
and subsequent bins span one magnitude to a lower limit of
 $M_{B}=-16.5$, close to the resolution limit. This 
results in 7 bins labelled \GL{1} to \GL{7} (Galaxy Luminosity)
from the faintest to the brightest. The limits of the
luminosity bins, and corresponding number of objects,
 are given in the fourth and fifth lines of Table~\ref{tab:BinLimits}.

\begin{table*}
\begin{minipage}{180mm}
\centering
\caption{The \emph{lower limits} of the bins used to define the halo and
galaxy samples. The last column gives the \emph{upper limit} of the 7th
bin. The first three rows are for the dark matter haloes binned by 
mass, the following pairs of rows are for the galaxies split by
luminosity ($M_{B}$), colour ($B-V$), 
and morphology ($M_{\rmn{B,bulge}}-M_{\rmn{B,total}}$), respectively.}
\begin{tabular}{@{}lllllllll@{}}
\hline
\rule{0in}{3ex}
Property & 1 & 2 & 3 & 4 & 5 & 6 & 7 & Upper limit\\

\hhline{=========}
\rule{0in}{3ex}
Halo particle number & 10 & 40 & 160 & 640 & 2560 &
10240 & 40960 & 163840 \\

\hline
\rule{0in}{3ex}

Halo mass ($M_{\odot}$)  & $5.1\times 10^{10}$  & $2.0\times 10^{11}$  &
$8.2\times 10^{11}$  & $3.3\times 10^{12}$ & $1.3\times 10^{13}$ &
$5.2\times 10^{13}$ &  $2.1\times 10^{14}$ & $8.4\times 10^{14}$ \\

\hline
\rule{0in}{3ex}
Number of objects & 183552 & 40871 & 11243 & 2948 & 861 & 214 & 38 & \\

\hhline{=========}
\rule{0in}{3ex}
Galaxy luminosity ($M_{B}$) & -16.5 & -17.5 & -18.5 & -19.5 & -20.5 &
-21.5 & -22.5 & $-\infty$ \\

\hline
\rule{0in}{3ex}
Number of objects & 120261 & 46473 & 17961 & 7103 & 2876 & 446 & 29 & \\

\hhline{=========}
\rule{0in}{3ex}
Galaxy colour ($B-V$) & 0 & 0.5 & 0.6 & 0.7 & 0.8 & 0.9 & 1.0 & 1.5 \\

\hline
\rule{0in}{3ex}
Number of objects & 16141 & 32801 & 78908 & 36193 & 12634 & 34759 &
8434 & \\

\hhline{=========}
\rule{0in}{3ex}
Galaxy morphology & $\infty$ & 4 &
2 & 1 & 0.75 & 0.5 & 0.25 & 0 \\

\hline
\rule{0in}{3ex}
Number of objects & 4255 & 7004 & 8219 & 3093 & 3410 & 1688 & 1748 & \\

\hline
\end{tabular}
\label{tab:BinLimits}
\end{minipage}
\end{table*}

\subsection{Colour selection}
\label{sec:Samples:Col}

From figure~7 of M01, we can infer that $B-V\sim0.8$ splits  
galaxies brighter than the morphological resolution roughly into ellipticals
and spirals. However, this split is
not perfect, as noted above for the reference galaxies. Moreover,
since we want to highlight a progressive 
change in the spatial distribution of galaxies with
a given property, we will proceed as in the previous paragraph. We split
the range covered by the $B-V$ index of all galaxies above the 
luminosity resolution limit of the simulation
($0\la B-V \la 1.42$ \footnote{$B-V \sim 1.4$ is the maximum reached
in our simulation: although stars are assumed to always form with
solar metallicity and a Scalo IMF, the approximate model for dust 
reddening that we subsequently apply to the mock galaxy catalogs 
yields a few galaxies that are very red}) in a series of 7 bins called 
\GC{1} to \GC{7} (Galaxy Colour), from the bluest to the reddest. 
The bluest bin is [0 0.5], and the following bin 
thresholds are separated by 0.1 in $B-V$. The last bin covers the
range [1 1.5]. Again, the limits of the bins, and number of objects, 
are given in the sixth and seventh lines of Table~\ref{tab:BinLimits}.

We stress that we have binned up all galaxies above the luminosity
resolution limit of our simulation. This allows us to go substantially
fainter than if we restricted ourselves to galaxies for which the
simulation gives reliable morphologies. 

\subsection{Morphology selection}
\label{sec:Samples:Morp}

Our last sample selection makes use of the morphogical information in
the simulation. By definition, we need here to restrict ourselves to
the galaxies brighter than the morphology resolution limit ($M_{B}\leq
-18.46$). The sample selection is done according to $M_{\rmn{B,bulge}}-M_{\rmn{B,total}}$
(which ranges from 0 to $\sim 7$ among the simulated galaxies which possess a
bulge). The bins are called \GM{1} to \GM{7} (Galaxy Morphology) from the least
bulge-dominated to the most bulge-dominated. Note that for simplicity,
we have put bulge-less galaxies (above the morphology resolution)
together with the objects in the first bin. The last two 
lines of Table~\ref{tab:BinLimits} give the limits 
of these ``morphology'' bins, together with the associated 
number of objects.


\section[]{The environments of the galaxy populations}
\label{sec:Envi}

To compare the local environments of the various samples, and
to look for a possible simple signature of a void population, we
first compute the mass density (by smoothing the DM field) of 
the region surrounding each galaxy, and we
compare the distributions of this local density in our various samples. In
addition, we estimate the luminosity function (per unit mass) of
objects lying in environments of given density, focussing particularly
on the variation of the overall shape.


\subsection[]{Density of the environments}
\label{sec:Envi:Dens}

\subsubsection{Method}
\label{sec:Envi:Dens:Method}

We characterize the environments of the galaxies by mass densities
smoothed over a 5 \hMpc smoothing scale (a smoothing of 10 \hMpc 
is already too large to bring out any trends between the samples). 

We assign the DM distribution of the simulation on a 
fine regular grid with mesh spacing
much smaller than one smoothing length ($R_{s}$), using a CIC scheme. 
The smoothing of the density field is performed on this fine grid 
 by means of a Gaussian kernel which takes the form:

\begin{equation}
	\label{eqn:SmoothKernel}
	W(r)={{1}\over{(2 \pi R_{s}^{2})^{3/2}}}\exp{\left({-{r^{2}}\over{2R_{s}^{2}}}\right)}
\end{equation}
where we take $R_{s}=5\:\hMpcDot$ 
Because we expect to study preferentially 
large voids with a diameter of $\sim 10\:\hMpcCom$ 
this scale seems appropriate.

Note that because we have only followed galaxy formation in the central, roughly
spherical, high-resolution region of simulation (with radius $\sim8000\:\kmsKC$ 
we also use a shell of low-resolution particles immediately
beyond to get the correct estimate of the DM density at mesh points
near the boundary. Once we have sampled the smoothed  DM field, we
then simply interpolate the overdensities computed 
on the grid to the positions of the DM haloes (given by their most bound
particle) or of the galaxies.

We consider then the normalized cumulative counts of the number of
galaxies (the population fraction) above a given mass overdensity threshold, as a function of
decreasing overdensity, starting from $\delta_{s}\ga30$ the maximum we
find for the $5\:\hMpc$ smoothing
length we use. If the galaxies of some test population reside 
preferentially in low-density environments, 
this will appear as a late
rise of the cumulative fraction 
with decreasing DM overdensity, compared, for instance, to the
behaviour of the reference galaxies.

By construction, the cumulative plots obtained from the galaxy
positions are mass- rather than volume-weighted. The
visual impression from the pictures of K99 recalled by P01 is one of very few 
simulated galaxies in the voids, with the latter filling a substantial
fraction of space. 
A simple way to assess the departure of the distribution of 
galaxies from a homogenous one with the tools of this Section is to 
use the regular mesh from which we have interpolated the DM density to
the galaxy positions. In the four plots of Fig~\ref{fig:Cumul5MpcFig}, the
repeated dotted line gives the cumulative fraction of mesh points
above a given smoothed DM overdensity threshold: it should be viewed
as the simulation volume fraction above the threshold. We will denote this
``mesh sample'' with \VCom and note that half of the simulation volume has a
DM environment density of $\delta_{s}$ below -0.24 and only about a
third of it has higher than average density.

In the same four plots, we also repeat the cumulative fraction of the
reference galaxies with a solid line. This line is very close to that 
we find for the DM particles themselves (thus for the ``mass'' in the
simulation) and is a translation by
almost a factor 2 towards higher density from the \V cumulative
fraction: each population fraction is reached in the \refg sample at
twice the DM density needed to reach the same fraction for the
uniformly distributed \V population.

\subsubsection{Results for mass and luminosity splitting}
\label{sec:Envi:Dens:ResultsMassLum}

In the top left panel of Fig.~\ref{fig:Cumul5MpcFig} we show with dashed lines the normalized cumulative
counts of the halo samples selected by mass (\HM{i}). In each of the four
plots of Fig.~\ref{fig:Cumul5MpcFig} that we discuss here, the lowest
and highest sample indices ($i=1$ and $i=7$) correspond to the
leftmost and rightmost dashed curves, respectively, with a monotonic
variation for the samples in between.

The first three halo samples \HM{1} to \HM{3} contain haloes with \emph{total} masses $M_{\rmn{tot}}\la
M_{*}$, where as usual we define $M_{*}$ such that
$\sigma(M_{*})=\delta_{\rmn{crit}}\sim1.69$  
($M_{*}\sim 1.44\times 10^{13} \msun$ here). The behaviours of the
fractions of cumulative counts for these three halo samples are very
similar: at overdensities under $\delta_{\tx{s}}\sim0.6$ they depart slightly
 from the counts of the \refg sample, favoring lower
density environments. However, the plots are still far from
the cumulative fraction for the grid counts; the population of haloes
is 90\% complete for $\delta_{\tx{s}}\geq-0.3$, while a third of the
simulation volume is at such low densities. 

While the plot of \HM{4} almost coincides with the \refg
cumulative fraction (88\% of the \refg
galaxies are central galaxies of haloes, and 53\% of them are
central galaxies of \HM{4} haloes), the three most massive halo bins
separate strongly from each other and from the \refg fraction. This is
a clear effect of the non-linearity of halo bias: 
according to the model of \citet{Mo96a}, one writes: 
\begin{equation}
b_{\rmn{halos}}(\nu)=1+\frac{\nu^2-1}{\delta_{\rmn{crit}}}\:,
\label{eqn:bias}
\end{equation}
where $\nu=\delta_{\rmn{crit}}/\sigma(M)$. Haloes less 
massive than $M_{*}$ are antibiased, haloes close to
$M_{*}$ like \HM{4} are unbiased, and the bias becomes more and more
substantial as one further increases the mass. Haloes of the most
massive sample completely avoid low and mean density regions (recall
also that the particles of a given halo contribute to the smoothed DM
density estimation of the region it resides in). In particular, there
are no haloes in the \HM{7} sample lying in environments with $\delta_{\tx{s}}\leq1.5$.

The top right panel of Fig.~\ref{fig:Cumul5MpcFig} shows the cumulative
fraction of the galaxy samples binned by their luminosity
(\GL{i}). The trends are similar to those of the halo samples selected
by their total DM mass, but except for the most luminous sample, 
the range of variation of galaxy bias is
much reduced. Of course this is due
to the fact that there is no tight relation between a halo mass and
the luminosity of its central galaxy, and that a given galaxy luminosity sample
has contributions from several differing halo samples. An
example of this dilution is the cumulative fraction of the counts of
the first four galaxy samples, which are very similar, and are 
very close to the \refg sample. The range of \Bband luminosities of the \refg
sample encompasses those of the \GL{4} and \GL{5} samples together. The
further match with the \GL{1} to \GL{3} samples shows that
the population of faint galaxies (i.e. taken globally in the samples 
and selected solely by luminosity) does not constitute 
a void population. \GL{4} and \GL{5} 
galaxies increasingly tend to avoid underdense regions 
($\delta_{\tx{s}}\leq0$) where a few \refg galaxies are found. 
The most luminous bin, \GL{7}, with only 29 galaxies, contains of
course a fair fraction of galaxies in very dense environments (BCGs in
regions with $\delta_{\tx{s}}\sim10$). However, as compared to \HM{7},
there \emph{are} also very bright galaxies in the \Bband in mean density
regions: these are galaxies which have undergone a recent merger
and are currently starbursting: among the 7 galaxies of the \GL{7}
sample with $\delta_{\tx{s}}\leq0.5$, 5 have a colour index
$B-V\leq0.5$ and $M_{B}\sim-23$. Again, this underlines the difficulty
in relating halo mass to central galaxy luminosity.

Recall that the small amount of (anti-)bias found when considering
faint luminosities or galaxies close to $L_{*}$ and the rather limited
positive bias for galaxies somewhat brighter (when not extremely
bright, though) is consistent with that found in the figure 14 of K99 
and with the correlation functions given below.

\subsubsection{Results for colour and morphology splitting}
\label{sec:Envi:Dens:ResultsColMorph}

\begin{figure*}
\begin{minipage}{180mm}
\centering
\epsfig{file=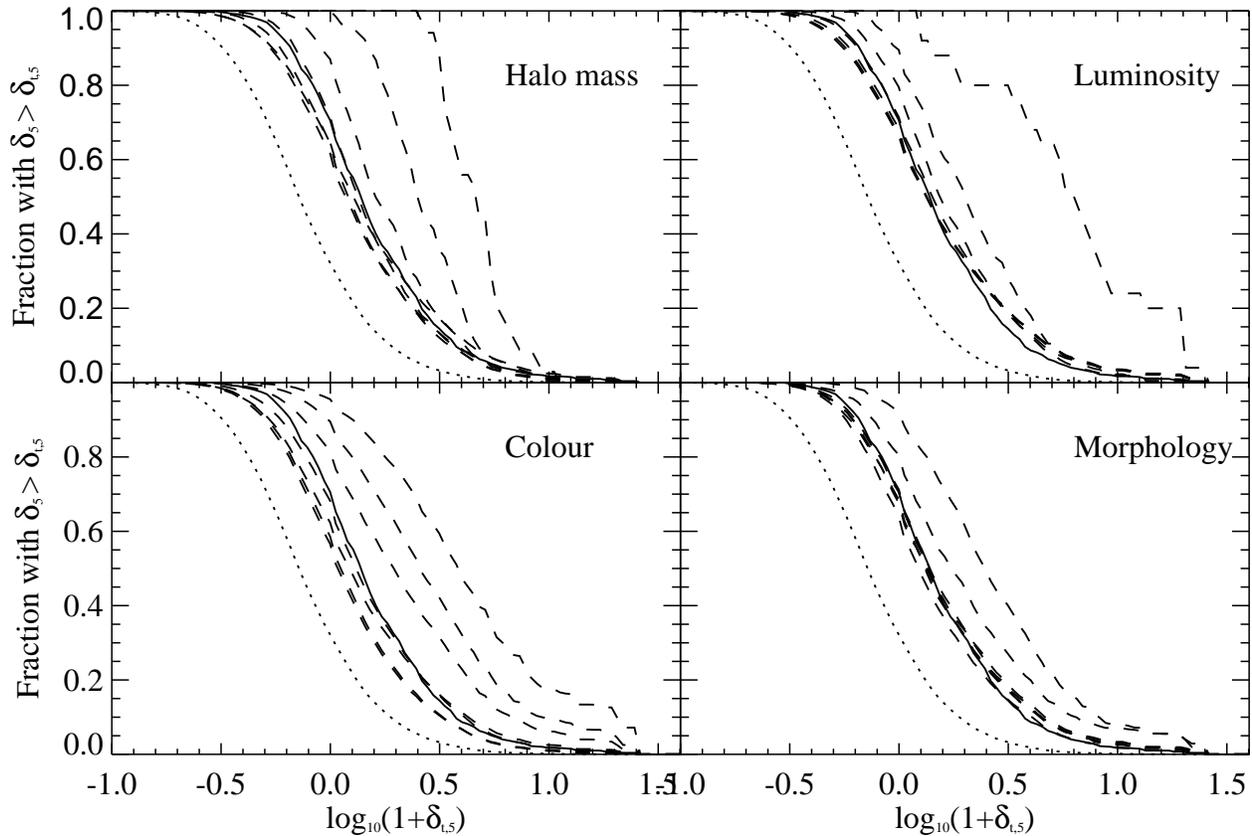,width=18cm,height=12cm}
\caption{The cumulative fraction of haloes and galaxies as a function
of their DM environment density smoothed over $5\:\hMpc$. In each
plot, the dashed lines give the cumulative fraction of the objects in
each of the samples (with index 1 and 7 associated to the leftmost and
rightmost curves respectively and monotonic index increase in
between). The repeated solid and dotted lines give the cumulative
counts for the \refg (reference spiral galaxies) and the \V (regular
mesh points) samples, respectively. The top left panel is for the
DM haloes split by mass (\HM{i}), the top right panel for galaxies
beyond the luminosity resolution limit, split by luminosity
(\GL{i}), and the bottom left and right panels give the cumulative
fraction for galaxies split by their colour (beyond the luminosity
resolution limit) and by their morphology (beyond the morphology
resolution limit), respectively.}																	
\label{fig:Cumul5MpcFig}
\end{minipage}
\end{figure*}

The bottom left panel of Fig.~\ref{fig:Cumul5MpcFig} plots the
 fraction of galaxies as a function of environment density among the
 colour samples (\GC{i}). Again, the different samples are
differently biased with respect to the \refg sample. Note, however,
 that a fair fraction of the galaxies in the two bluest samples populate
 regions more underdense than any of the other galaxy 
 samples considered here: the curves of \GC{1} and \GC{2} 
are close to each other and rise later
 than that of the \refg sample (the \HM{1} halo sample also ``favours''
 such underdense regions as compared to the usual trends).  
10\% of the \refg population lies in
 environments with densities $\delta_{\tx{s}}\leq-0.37$ while 10\% of
 the \GC{1} population lies at $\delta_{\tx{s}}\leq-0.5$. Of course,
 this is still substantially denser than the corresponding density threshold for
 the \V sample: such blue galaxies do not make a
 homogeneous galaxy population on their own, but they nevertheless
 constitute the most promising ``void fillers''. 
For increasingly redder populations, a preference for denser
 environments develops, but even in the reddest sample, there
is a small fraction of the population in underdense regions, quite
 similarly to the \GL{7} sample. The fraction of the reddest 
galaxies in very dense regions ($\delta_{\tx{s}}\sim10$) is of order of 15\% and
  smaller than the 30\% fraction for the \GL{7} sample (recall also
 that there are 8434 galaxies in \GC{7}, and only 29 in \GL{7}). 
Leaving aside\ \GL{7}, one concludes that the variation of bias is
 stronger among colour selected samples than among luminosity selected
 samples.

The bottom right plot of Fig.~\ref{fig:Cumul5MpcFig} deals with
our morphology selected populations (\GM{i}). The five samples with later type
galaxies show plots of cumulative fractions quite close to those of
the reference spirals. Bias with respect to the \refg sample,
favouring denser galaxy environment, is visible for the two bulge
dominated samples. Yet, the preference towards higher densities for
extreme \GM{7} galaxies is weaker than in our other extreme samples.

To conclude, none of our \GM{i} populations has galaxies 
filling low density regions. Note that the morphology--density 
relation that we have implicitly constructed is limited 
to objects with luminosities \emph{above} the 
morphology resolution at $M_{B}\sim-18.5$. This might partly explains why we
found the \GC{1} galaxies (down to the luminosity resolution) 
to reside on average in lower density regions than the \GM{1}
galaxies. Low mass haloes with less than 100 particles are
generally not expected to host galaxies with $M_{B}\la-18.5$, simply
because they are not massive enough. Then, one can easily understand
why any sample constructed by splitting the $M_{B}<-18.5$ galaxy
population will not show a curve of cumulative counts below that say,
of the \HM{3} halo sample.


\subsection[]{Luminosity functions}
\label{sec:Envi:LF}

The previous section has shown that the typical environment 
(when smoothed on a $5\:\hMpc$ scale) of the galaxies varies only
slowly with their luminosities. Conversely, one can look 
for the distribution of galaxy luminosities as a
function of the surrounding DM overdensity. Because a given
Eulerian volume corresponds to different mean DM densities and
different mean number of galaxies, when we probe
different environments, we split our discussion of the luminosity
functions (LF) in two series of five adjacent DM environment density bins. The first
series (called \LFV{i}) is set so as to have the same simulation
\emph{volume} in each bin (hence probing 20\% of the whole volume), and the second
one (called \LFM{i}) to have the same \emph{mass} in each bin
(20\% of the whole mass). In each of the two cases, 
we will express the luminosity function in units of
counts per magnitude. Obviously with the appropriate \emph{same} vertical shift
of all LFs of the \LFV{i} sample, we could get units of counts per
magnitude per unit volume, and with another \emph{same} shift
for all LFs of the \LFM{i} sample, we could get units of counts per
magnitude per unit DM mass. In the first line of 
Table~\ref{tab:LumFunDensThres} we give the five
lower limits for the DM environment density bins (the upper limit of 
the last bin is left open), for the \LFV{i} and \LFM{i} samples 
(left and right columns respectively). 
In the second and third lines of the table, we give the
corresponding minimum \Bband magnitude in each bin, and the number 
of contributing galaxies. 

Here, we are mostly interested in the variation of the 
shape and normalization of the luminosity function with
environment. For instance, if faint dwarf galaxies are to avoid high-density
regions and to populate preferentially the voids (for a given enclosed
DM mass), one would expect a steeper slope in lower density
regions. The environments are again characterized 
by the DM overdensity smoothed with
a gaussian kernel of dispersion $5\:\hMpcDot$

\begin{table*}
\caption{Minimum smoothed DM density thresholds ($\delta_{s}$)
defining the bins where the \LFV{i} (left column) and
\LFM{i} (right column) luminosity functions are computed. 
The second line gives the minimum $B$-band magnitude among 
galaxies in each interval, and the third line the number of
contributing galaxies.}

\begin{tabular}{@{}cc@{}}

Equal volume (\LFV{i}) & Equal mass (\LFM{i}) \\

\begin{tabular}{@{}llllll@{}}

\hline
\rule{0in}{3ex}
Bin number & 1 & 2 & 3 & 4 & 5 \\
\hline
\rule{0in}{3ex}
Minimum density  & -1  & -0.57 & -0.38  & -0.14  & 0.33  \\
\hline
\rule{0in}{3ex}
Minimum $B$-band magnitude   & -20.9 & -21.5 & -23.1 & -23.1 & -23.6 \\
\hline
\rule{0in}{3ex}
Number of galaxies  & 5215 & 13179 & 20761 & 33059 & 89242 \\
\hline
\end{tabular}

&

\begin{tabular}{@{}lllll@{}}

\hline
\rule{0in}{3ex}
1 & 2 & 3 & 4 & 5  \\
\hline
\rule{0in}{3ex}
-1  & -0.23 & 0.23  & 1 & 2.6 \\
\hline
\rule{0in}{3ex}
-23.1 & -23.1 & -23.0 & -23.8 & -23.6 \\
\hline
\rule{0in}{3ex}
32068 & 34025 & 34940 & 29723 & 30700 \\
\hline
\end{tabular}

\\

\end{tabular}
\label{tab:LumFunDensThres}
\end{table*}

\begin{figure*}					
\begin{minipage}{180mm}
\centering
\epsfig{file=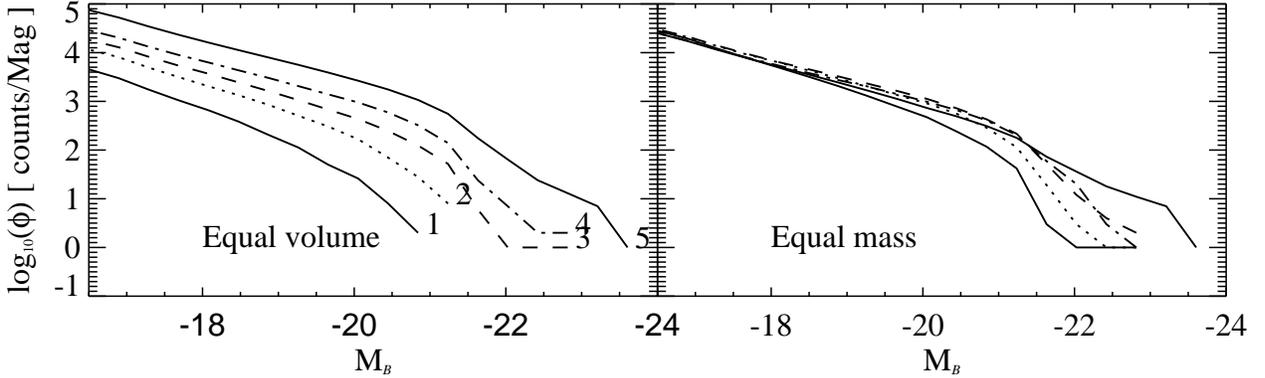,width=18cm,height=6cm}	
\caption{The \Bband luminosity functions (in units of
galaxy counts per magnitude) for the 5 \LFV{i} and \LFM{i} bins
(left and right panels respectively). Each of the \LFV{i} and \LFM{i}
bins covers crespectively the same 20\% simulation 
volume and contains the same 20\% of the total simulation DM mass. The tags 
at the bright ends of the LFs in the left panel refer to the bin indices as given by the first line of
Table~\ref{tab:LumFunDensThres}. The coding of the lines has no special
meaning, but alternates for clarity. It is the same in both panels. }
\label{fig:LumFunEnvFig}
\end{minipage}
\end{figure*}

The left and right panels of Fig.~\ref{fig:LumFunEnvFig} give the series of 
\Bband LFs, for the \LFV{i} and \LFM{i} samples respectively. 
 The tags at the bright ends of the functions label the
id's of the different environments (first line of
Table~\ref{tab:LumFunDensThres}). The coding of the lines has been
alternated for clarity.


These LFs call for several comments, dealing
 successively with the minimum magnitude reached by the selected
 galaxies, the normalisation of the overall LF, the exponential
 cutoff and the faint-end slope. 

\begin{itemize}

\item The left panel of Fig.~\ref{fig:LumFunEnvFig} shows how LFs
computed for two most underdense bins in the \LFV{i} sample do not reach as
high luminosities as the LFs computed for higher density bins. The
second line of Table~\ref{tab:LumFunDensThres} gives the minimum  \Bband
magnitude of the galaxies in each bin. While they hardly
reach $M_{B}\sim-21$ and $M_{B}\sim-21.5$ for
$\delta_{\tx{s}}\la$ -0.57 and -0.38 respectively, there are already
bright galaxies with \Bband luminosities characteristic of the BCGs of massive
clusters ($M_{B} \sim -23$) in the third 
and fourth bins with $-0.38\leq\delta_{\tx{s}}\leq0.33$,
probing volumes with DM densities close to the mean. These brightest
objects are typically  undergoing a starburst associated 
with a major merger, which brighten up in the \Bband. Recall that our
adopted TF normalization means that, on average, a central spiral
galaxy of a halo with $\Vvir \sim 220\:\kms$ has $M_{B}\sim -20.8$.
Of course, the last sample \LFV{5} contains the BCGs of the most massive clusters, and
as expected we find there the brightest galaxy ($M_{B} \sim -23.6$) of the whole
simulation. The first four LFs of the \LFM{i} bins given in the right 
panel of Fig.~\ref{fig:LumFunEnvFig} have comparable maximum galaxy
luminosities. The absence of segregation from the point of view of
the luminosity of the brightest galaxies among the bins of this sample
is only due to the modified density thresholds of \LFM{i} 
compared to those of \LFV{i}: already \LFM{1} probes smooth DM
densities reaching high enough to include the starbursting
galaxies mentioned above.

\item Although the low density regions covered by \LFV{1} constitute
20\% of the simulation volume, they only contain 4\% of the total
mass. At the other extreme, the \LFV{5} bin carries 57\% of the
total mass in quantitatively the same volume 
and naturally hosts a larger number of galaxies. 
This readily explains the difference in the overall
normalization (counts of galaxies) of the LFs in the \LFV{i} samples, 
as one goes from low density regions to high density ones. As 
expected, the LFs in the \LFM{i} sample have quite similar
normalizations (at $M_{B}\sim-17$): each bin contains a similar mass. 
Interestingly, even the first bin which is restricted to $\delta_{\tx{s}}\leq-0.23$ 
and which encompasses 55\% of the simulation volume does not host a
very different number of faint (say with $-16.5>M_{B}>-17.5$) galaxies
per DM mass than does, e.g. \LFM{3} which covers 13\% of the volume. This argues
against a strong luminosity--density relation at the faint end (at the
simulation resolution) and seems to confirm that the faint galaxy population
does not predominantly reside in voids.

\item An ``exponential'' cutoff is visible for the samples \LFV{3} and
\LFV{4} probing mean density environments. It is weaker for
\LFV{1} and \LFV{2}, simply because in ``void'' and underdense regions, 
the LFs do not reach bright enough luminosities 
where one would expect the cutoff. The most overdense
sample \LFV{5} shows a complicated behaviour, one can 
notice a possible cutoff at $M_{B}\sim-21$, beyond which 
the LF exhibits sort of an ``inflexion point'' at $M_{B}\sim$-21.5 or
-22. This presumably echoes the difficulties encountered by M01 
in obtaining good-looking cluster luminosity functions 
at the bright end. Again due to the shift in
maximum density thresholds towards higher densities for all \LFM{i}
bins compared to the \LFV{i} sample, cutoffs at $M_{B}\sim -21$ make a
clear feature of the first four samples of \LFM{i}. It is reassuring
for the galaxy formation scheme that these ``knees'' break 
the LFs at very similar magnitudes. Again,
the presence of a cutoff for \LFM{5} is less obvious because of the
rather uncertain modelling of the luminosity of the BCGs of massive clusters.

\item Faint-end slope variation is visible in both samples, and is more
striking in the case of \LFM{i}. Over the range [-16.5 -19.5], a rough
estimation is $\alpha\sim-1.6$ and $\alpha\sim-1.4$ 
in the lowest and the highest density bins respectively. 
These estimations repeat for \LFV{i}.  This steepening
in the faint-end slope in underdense regions 
is due to the relative depletion of $L_{*}$ 
galaxies in such environments when compared to mean density 
or overdense environments, while the 
counts of faint galaxies with $M_{B}\sim-17$ are similar in the
two cases. 

\end{itemize}

The above results assume that at the lowest luminosities considered, no
major physical process has been neglected. In particular, the
extrapolation to even fainter luminosities of the model for galaxy formation 
that has been used here is hazardous: for example, background UV
radiation \citep{Th96} must be included when computing the infall
rate of the cold gas on haloes with small mass ($\Vvir\sim30-50\:\kms$),
which collapse late, after reionization. 
Similarly, reheating by SNe feedback could very well expell
 a large fraction of the cold gas from the smallest haloes and suppress
subsequent star formation there. Note also, as discussed at length in
M01, that the overall luminosity function of the model analyzed here is a
relatively poor fit to that observed. The results of this Section
should thus be taken as indicating the expected trends in cutoff
luminosities and faint-end slopes rather than their specific values.


\section[]{Clustering of the galaxy populations}
\label{sec:Clus}

In this Section, we consider ``two-point'' statistics: first we check
the correlation functions (in real space) of our samples. 
We recover the usual results, and underline
 the particularly low clustering of the ``blue'' sample with respect
to the underlying DM, the latter being at the level of the reference
spiral population. We then use nearest
neighbour statistics to gain better insight into the relative spatial
distribution of the reference galaxies and the various test populations.


\subsection[]{Correlation functions}
\label{sec:Clus:Corr}

\begin{figure*}					
\begin{minipage}{180mm}
\centering
\epsfig{file=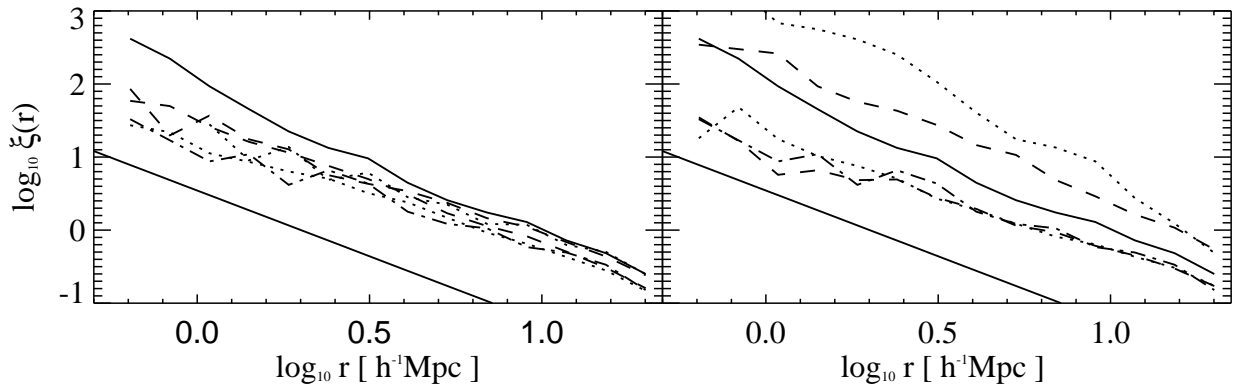,width=18cm,height=6cm}
\caption{The real-space correlation functions of our
samples. The left panel shows the correlation of the \HM{1} and \GL{1}
samples (dotted and dashed lines respectively) and of
the \HM{5} and \GL{5} samples (dotted and dashed lines, above the
two previous ones); the right panel gives
the correlation functions of the \GC{1} and \GC{7} samples (dotted lines, 
with \GC{1} below the solid line) and of
the \GM{1} and \GM{7} samples (dashed lines, with \GM{1} below the solid line). 
In both panels, the solid line is the correlation of the dark matter, 
and the dashed-dotted line that of the reference galaxies. The
straight solid line in the lower left corner of each panel
shows a -1.8 slope. }
\label{fig:CorrFuncFig}
\end{minipage}
\end{figure*}

In both panels of Fig.~\ref{fig:CorrFuncFig}, we
reproduce the real-space autocorrelation functions of the DM (upper solid
line), together with the autocorrelation functions 
of our reference galaxies (dash-dotted line). In the left panel, we
show the correlations of the \HM{1} and \HM{5} samples (dotted lines,
with \HM{1} having the lowest amplitude on large scales), 
and the correlations of the \GL{1} and \GL{5} samples (dashed lines,
with \GL{1} having the lowest amplitude on large scales). In the 
right panel, we give the correlations of the
\GC{1} and \GC{7} samples (dotted lines, with \GC{1} below the solid
line) and the correlations of the \GM{1} and \GM{7} samples (dashed
lines, with \GM{1} below the solid line). Note that we show the
correlations of \HM{5} and \GL{5} instead of \HM{7} and \GL{7} 
simply because of the small number of objects in these last two samples, which results
in too noisy correlation functions. The straight solid line in the
lower left corner of each panel shows a -1.8 slope. All correlation functions have
been computed from $500\:\hkpc$ to $20\:\hMpc$. When interpreting
these plots it is important to remember that the region simulated is
quite small and is constrained to match the large-scale structure of
the observed nearby galaxy distribution. As M01 show, the model
autocorrelation functions of the various  galaxy populations agree
well with the observed functions for galaxies in the corresponding
volume.

The logarithmic slope $\gamma$ of the autocorrelation of the dark
matter is close to -1.8 . In the left panel, all samples, including \refgCom
have a shallower slope, closer to -1.4 . This value is also
close to the slopes of both the \GC{1} and \GM{1} samples in the 
right panel. The other extreme samples, \GC{7} and \GM{7}, have steeper correlation
functions. The curves differ most noticeably, however, 
in amplitude (bias), even on large
scales ($\sim20\:\hMpc$). Note that the steeper slope of \GC{7}
compared to, e.g., \GC{1} is a consequence of the colour bias
being stronger at smaller scales, as is evident in the observational
sample of \citet{Wi98}.

In the left panel of Fig.~\ref{fig:CorrFuncFig}, it is striking 
how the autocorrelation of \HM{1} matches that of the
reference spirals, on scales greater than $\sim 2\:\hMpcDot$ The
autocorrelation of \GL{1} also approaches the correlation of the
\refg sample on these scales, but it is somewhat stronger. On
smaller scales, the \GL{1} sample is more correlated than both
\HM{1} and \refgDot On the other hand, the autocorrelations of \GL{5} and \HM{5} are very similar on all scales
shown. The reason might be that 88\% of the galaxies of the \GL{5} are
central galaxies of haloes, and 82\% of them belong to \HM{4} or \HM{5} haloes.  

Note that the bias between \GL{1} and \GL{5} is apparent on scales
greater than $4\:\hMpcCom$ beyond which it stays constant at a level
of 1.2 to 1.3. A similar behaviour for the halo bias is seen between \HM{1} and
\HM{5}. Such a low level of respective bias is surprising given the rather
different masses in the two samples: from equation~\ref{eqn:bias}, one
would expect $b_{1}=0.53$ and $b_{5}=1.14$ and so
$b_{51}=2.15$, using the mean masses of the haloes in each sample, 
 but note that the more accurate formula of
\citet{Ji98} predicts somewhat smaller effects in the sub-$M_{*}$
range ($b_{51}\sim1.7$). We 
have also already seen that the moderate 
amount of galaxy bias does not
contradict the one given in the top left plot of figure 14 of K99
(although for brighter galaxies).


Finally, the correlation of the reference galaxies is notably 
antibiased with respect to the DM on small scales ($\la 3\:\hMpc$). On
larger scales, the antibias reaches an almost constant value of
$1.2$. Again, this antibias of the late-type, star-forming 
galaxies was also noticed by K99.

As shown in the right panel of Fig.~\ref{fig:CorrFuncFig},
 the bias between \GC{1} and \GC{7} and between \GM{1} and \GM{7} 
is much stronger than between \GL{1} and \GL{5} for instance. 
As for the left-hand plot, the
correlation functions of the first samples \GC{1} and \GM{1} match
very well the autocorrelation of the reference galaxies. We therefore 
stress that the correlation of the bluest galaxies
is \emph{not} much weaker than that of the reference spirals. The
correlation of the elliptical galaxies (\GM{7}) parallels that of the DM
with a strong constant bias of $\sim1.6$ with respect to the
matter. Consistently with the analysis of K99, the colour bias is even
 higher than the morphology bias, but varies significantly with scale,
from $b\ga3$ on small scales of order $2\:\hMpcCom$ until it reaches
the smaller bias of the elliptical galaxies at $15\:\hMpc$. This is
can be explained by the contribution of faint, very red, satellite galaxies
in massive haloes which increases the correlation function on small
scales, while the morphology luminosity limit applied in the
elliptical sample \GM{7} suppresses such contribution.

This discussion of autocorrelation functions should be viewed 
as a mere check that we recover the usual trends among the samples. 
As stressed by P01, it is not possible to infer the relative
spatial distribution of the populations on the basis of their
correlation functions alone.

\subsection[]{Nearest Neighbour Statistics}
\label{sec:Clus:NN}

\subsubsection[]{Method}
\label{sec:Clus:NN:Method}

The nearest neighbour analysis compares the cumulative 
distribution of the distances
$D_{to}$ to the nearest ``ordinary''
neighbour from galaxies in a test sample 
(for example, irregulars) to the cumulative distribution
of the distances $D_{oo}$ from an ordinary galaxy to the nearest other
 ordinary galaxy. If an extended tail appears in the distribution of $D_{to}$ compared
to $D_{oo}$, the spatial distributions of ordinary and test galaxies
differ. In this case, if ordinary galaxies define large voids, test galaxies will
tend to populate them. The reader is redirected to P01 for more details.

Here, given the number density of the galaxies in our samples and the probable maximum
radius of the voids in the simulation, we expect that both the maximum
distance of a t-galaxy to its nearest o-galaxy and the
maximum distance of an o-galaxy to its nearest o-galaxy
will be of order 15 \hMpcDot As a result, we limit the centers (t and o)
 to the innermost 65 \hMpc around the MW (the high-resolution
zone of the simulation has a radius of $\sim 80\: \hMpc$). To avoid
possible complications due to peculiar velocities of galaxies, 
we compute nearest neighbour statistics in real
space. Although P01 carries out his whole analysis in redshift space,
we have checked that the induced difference is negligible for the points we
want to make. 

To assess the significance of our results (and again following P01), 
we also randomize in the whole simulation volume 
the positions of one third of the galaxies in each test sample, and
recompute $D_{to}$. This shows the effect of
having at least \emph{one third} of the test population \emph{homogeneously}
distributed. In the following, we consider this level as our lower
``detection'' limit for finding a void population. This
procedure is clearly somewhat arbitrary but allows direct 
comparison with P01. 

\subsubsection[]{Results}
\label{sec:Clus:NN:Results}

\begin{figure*}					
\begin{minipage}{180mm}
\centering
	\epsfig{file=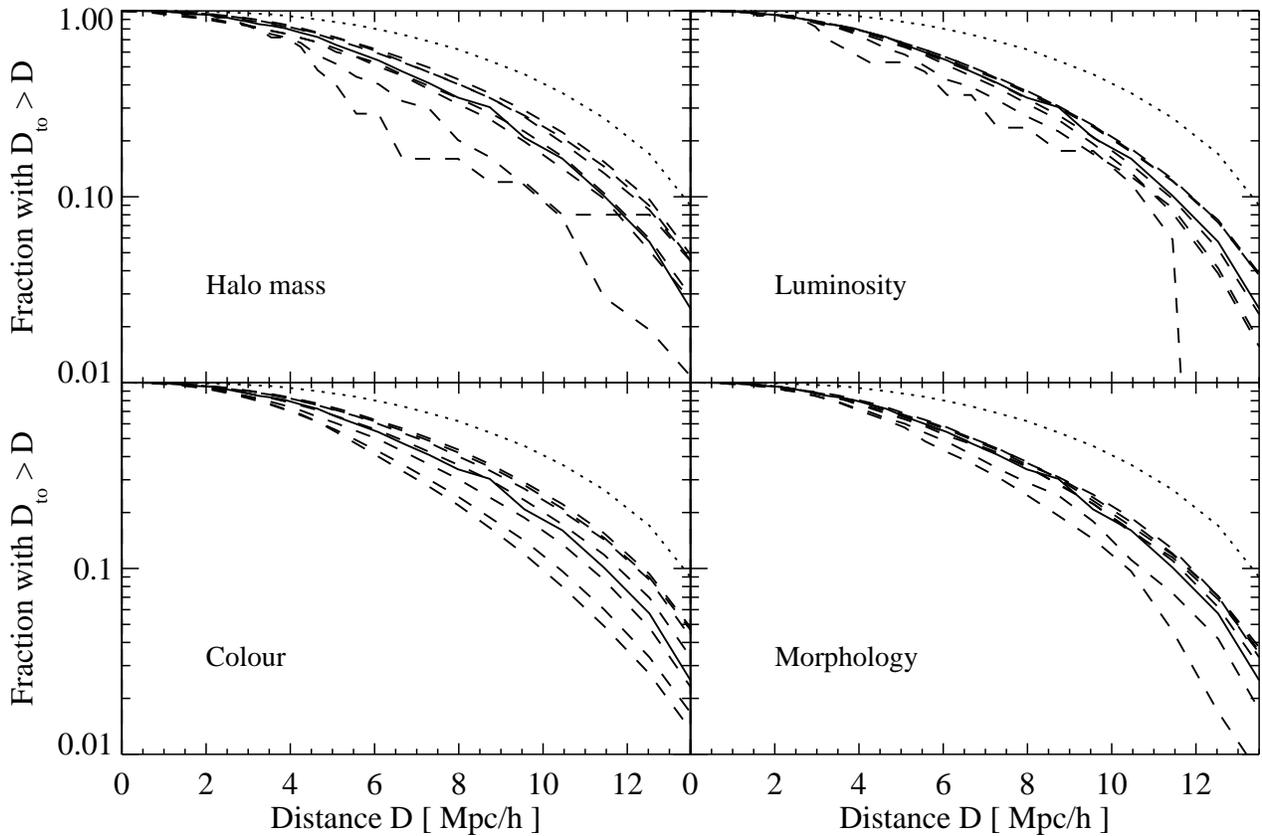,width=18cm,height=12cm}	
	\caption{The cumulative distribution of the distances $D_{to}$
	from the test galaxies of the various samples to the nearest reference galaxy (dashed lines), 
	the cumulative distribution of the distances $D_{oo}$ from
	reference galaxies to the nearest other reference galaxy (solid
	line), and the cumulative distribution of the distances $D_{to}$ 
	measured from the nodes of a regular mesh to their nearest
	reference galaxy (dotted line). The sample indices associated with the
	curves increase from the right to the left.}
	\label{fig:CNN_All_Fig}
\end{minipage}
\end{figure*}

Fig.~\ref{fig:CNN_All_Fig} gives the nearest
neighbour statistics for each of our four sets of samples (see labels). 
In each case, the cumulative fraction of the distance from
a reference galaxy to its nearest reference galaxy ($D_{oo}$) is shown by the
solid line (the same in all the plots in this Section). 
The cumulative fraction of the 
distance from a test galaxy of the various samples to
its nearest reference galaxy ($D_{to}$) is given by the dashed
lines. In all cases, the rightmost dashed curve corresponds
to index 1 of the samples, the leftmost dashed curve to 
index 7, and the indices increase monotonically in between. 
To show the level of a homogeneous population, we spanned the
innermost $65\:\hMpc$ of the simulation with a regular mesh with cell
size $5\:\hMpcCom$ resulting in $\sim10000$ points. We chose this 
size to end up with approximately the same density as in our
bluest sample of galaxies, \GC{1}. The cumulative fraction of such mesh 
points as a function of the distance to their nearest spiral neighbour
is repeated with dots in the four panels of the Figure.

\begin{figure*}					
\begin{minipage}{180mm}
\centering	
	\epsfig{file=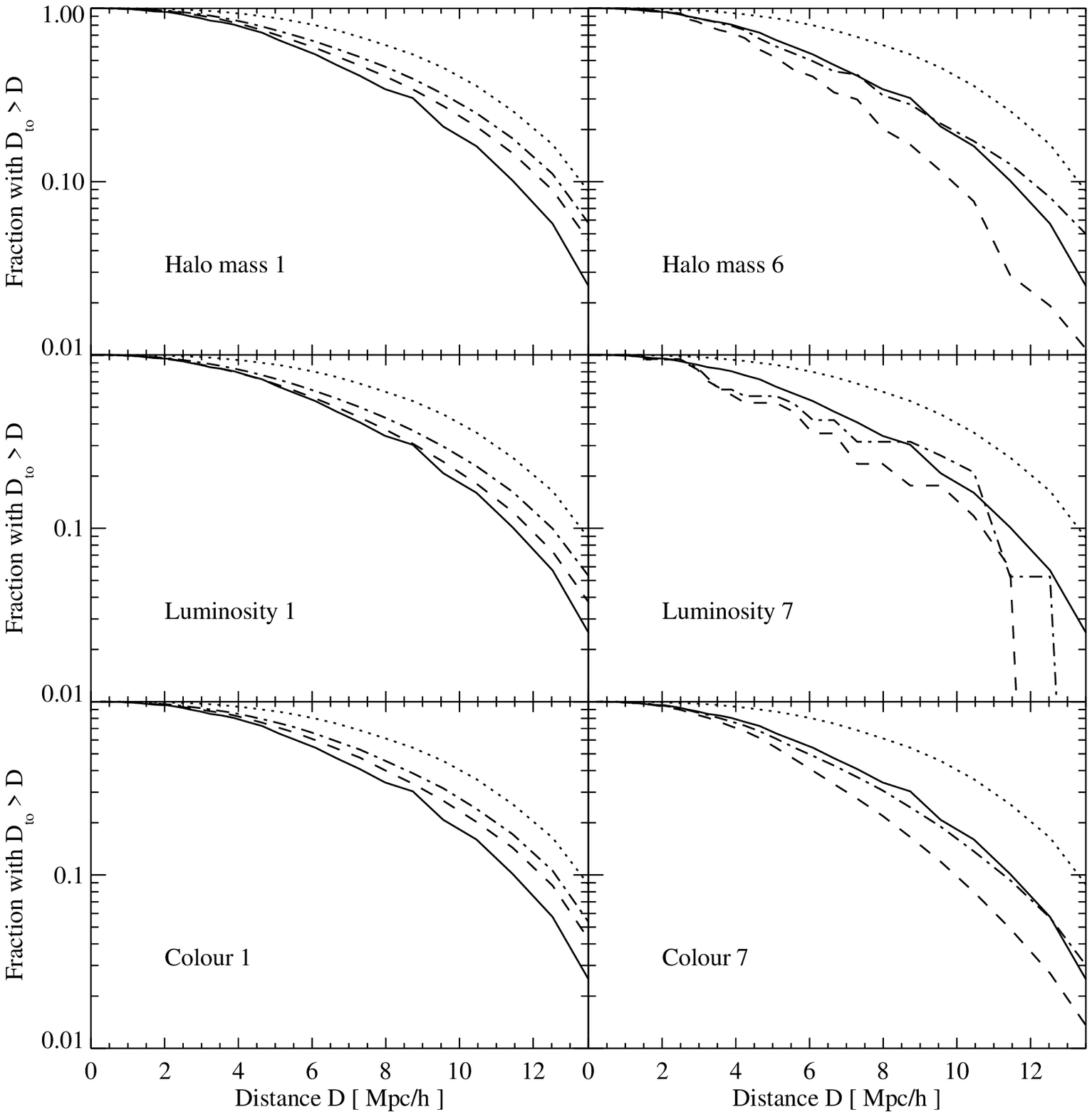,width=18cm,height=18cm}	
	\caption{The dashed line gives $D_{to}$ for the extreme bins
of each sample (1 and 7, with the exception of \HM{6}). 
The dash-dotted lines show the corresponding $D_{to}$ after randomizing the positions of 
one third of the objects in each test sample, and the solid curve
gives $D_{oo}$.  The dotted line is the same as in the previous Figure.}
	\label{fig:CNN_All_1_6_Fig}
\end{minipage}
\end{figure*}

The Fig.~\ref{fig:CNN_All_1_6_Fig} shows   
the cumulative fractions $D_{to}$ for the extreme bins (1 and 7) 
(as an exception, \HM{6} in the right panel for
reasons of noise) of each of our population samples, as computed
initially (dashed lines) and after randomizing the positions of one third of the test
galaxies through the whole simulation volume (dash-dotted lines). 
The cumulative fraction of the mesh 
points as a function of the distance to their nearest spiral neighbour
is also shown with dots. Note, as expected, that we found the curves of the
randomized objects of the intermediate samples 
always to lie between those of the randomized objects of the extreme
samples, and so we do not plot them.

 The signature of halo bias is well visible. The haloes of the least
massive sample, \HM{1}, have farther nearest spiral galaxies than do the spiral
galaxies themselves ($D_{to}>D_{oo}$): the rightmost dashed curve 
lies well above the solid curve. This holds for the next two
 samples too, but to a lesser extent. Hence, these small haloes 
are more broadly distributed than the reference spirals. 
The overall agreement of $D_{to}$ with $D_{oo}$ for \HM{4}  
 and for \HM{5} ($\Mvir\sim M_{*}$) can be 
understood in the same way as in Fig.~\ref{fig:Cumul5MpcFig}.

At separations greater than $\sim4\:\hMpcCom$ 
the two most massive halo samples 
have nearest spiral neighbours which are on average 
much closer than those of the \refg galaxies. Of course, such massive 
haloes are found almost exclusively in dense
environments with more galaxies, and can also be in the process of
accreting spirals from the field. Also recall that it is possible for
haloes to find their nearest neighbour among their own galaxy population.

 Only a fraction of the least massive haloes  
 \emph{could} be homogeneously distributed with respect to the
spirals, but in the top left panel of Fig.~\ref{fig:CNN_All_1_6_Fig}, 
 the dash-dotted line of the cumulative fraction of $D_{to}$  
for the \HM{1} sample after randomization 
of the positions is above the dashed line.
Less than a third of the low mass haloes are homogeneously
 distributed. The difference between the dash-dotted and dashed
line is much larger for the \HM{6} haloes in the right
panel, because they are more clustered, and the randomization thus has
 a proportionally greater effect.

The trend goes in the same direction, but is weaker, for the \GL{i} samples of galaxies
selected by luminosities. There is only limited difference in the
distribution of these samples of galaxies with respect to the
reference spirals. Position randomization acts in the
same way as for the haloes, but the cumulative fraction of the
partly randomized test objects is always the highest. Less 
than a third of even the faintest galaxies 
is homogeneously distributed.

The colour-selected, blue \GC{1} sample would be the best candidate for a
  population filling the voids between spirals.  However,
 after partly redistributing its population, the distance to the nearest spiral 
at a given fraction of the objects is still higher than for the
initial \GC{1} sample. Although the difference between true and randomized 
samples is somewhat weaker than in the previous cases, the conclusion
 for a homogeneous population is negative. Note that galaxies in \GC{1} also have the
highest average SFR per stellar mass (among all colour selected
samples), and may be termed ``active galaxies''.  
Conversely, the reddest sample has nearest spiral neighbours 
clearly closer than have the spirals themselves, consistent with the
 above results for massive clusters, and the fact that very red galaxies are
mostly satellite galaxies of such systems.
  
The trends in nearest neighbour statistics for the morphologically selected
samples (\GM{i}) are similar to the results obtained for colour
selection, but with a somewhat lower amplitude.

A detailed comparison with the observational results presented in P01
is difficult both because the analysis there is carried out in
redshift space and, more importantly, because the definition and
completeness of the observational samples are difficult to
quantify. The qualitative agreement is, however, quite good. The most
broadly distributed subsamples in our simulation (e.g. \GC{1}, GL{1} or
\GM{1}) have nearest neighbour distributions which relate to those of
the reference spirals in much the same way as P01 finds for his
observed samples of dwarf and LSB galaxies. In addition, the change in
the distributions caused by randomizing the positions of a third of
the test galaxies, are similar in the simulated and observed
samples. We conclude that the nearest neighbour statistics suggest
that the behaviour of the observed and simulated populations with
respect to voids are quite similar.

\section[]{Conclusions}
\label{sec:CCL}

We have looked for signatures of voids in the \lcdm simulation of M01, which mimics the dark matter
and galaxy distribution of the Local Universe up to 8000 \kmsDot The
simulation can resolve the morphology of an  LMC-type galaxy and the
luminosity of a dwarf elliptical. We have 
 addressed the question, raised by P01, 
of whether numerical simulations of galaxy
formation in the current standard picture predict objects 
in the observationally empty spaces defined by normal, $L_{*}$ spirals. 

We showed first that regions of size $\sim10\:\hMpc$ exist in the simulation 
which are devoid of even the smallest DM haloes we can resolve. 
We studied the distribution of galaxies as a function of 
luminosity, colour and morphology, and the halo
distribution as a function of mass. We found that \emph{none} of our samples
fills in underdense DM environments. The faint-end slope
of the ``equal-mass'' luminosity functions computed in regions 
of different densities shows some steepening as one goes to less dense
regions: dwarfs are relatively more abundant than $L_{*}$ galaxies
in comparison with high density regions, but the overall
variation of the shape of the LF is limited. Nearest 
neighbour statistics suggest that none of our
simulated populations can be considered to fill 
in the voids defined by $L_{*}$ spirals. This 
contrasts with the discussion of P01, who states that 
 at z=0 there is still a significant fraction of the matter in regions
between clusters and filaments in simulations of a
flat, low-density universe. Down to its resolution limit, our scheme of
galaxy formation qualitatively reproduces the observed galaxy
populations around voids.

The present simple study can be
expanded in two ways. First, one might derive a more quantitative
comparison based, for instance, on the distribution of the sizes of the voids,
which could be compared to the observations of \citet{Mu00} or
to the analytical model of \citet{Fr00}. Second, 
one can go to higher resolution. However, with 
current computer capacities, simulations of the size of the one
exploited here are already costly. Furthermore, a comprehensive reevaluation of
the relative importance of the physical processes would be needed. For
example, nearby ionizing sources at $z\sim3$ or the general 
UV background may inhibit the formation of galaxies like the Fornax dwarf 
in a spatially modulated way.


\section*{Acknowledgements}

The simulations presented in this paper were carried out on the T3E
supercomputer at the Computing Center of the Max-Planck-Society in
Garching, Germany. 

Simulated galaxy populations analysed here are publically available at \\ 
http://www.mpa-garching.mpg.de/NumCos/CR/
 
\bsp

\label{lastpage}

\bibliographystyle{mnras}
\bibliography{Voids251101}

\end{document}